# Rayleigh imaging in spectral mammography

Karl Berggren[*ab], Mats Danielsson[a] and Erik Fredenberg[b]

[a]Department of Physics, Royal Institute of Technology (KTH), 10691 Stockholm, Sweden
[b]Philips Healthcare, Smidesvägen 5, 17141 Solna, Sweden

**ABSTRACT**

Spectral imaging is the acquisition of multiple images of an object at different energy spectra. In mammography, dual-energy imaging (spectral imaging with two energy levels) has been investigated for several applications, in particular material decomposition, which allows for quantitative analysis of breast composition and quantitative contrast-enhanced imaging. Material decomposition with dual-energy imaging is based on the assumption that there are two dominant photon interaction effects that determine linear attenuation: the photoelectric effect and Compton scattering. This assumption limits the number of basis materials, i.e. the number of materials that are possible to differentiate between, to two. However, Rayleigh scattering may account for more than 10% of the linear attenuation in the mammography energy range. In this work, we show that a modified version of a scanning multi-slit spectral photon-counting mammography system is able to acquire three images at different spectra and can be used for triple-energy imaging. We further show that triple-energy imaging in combination with the efficient scatter rejection of the system enables measurement of Rayleigh scattering, which adds an additional energy dependency to the linear attenuation and enables material decomposition with three basis materials. Three available basis materials have the potential to improve virtually all applications of spectral imaging.

## 1. INTRODUCTION

Material decomposition with spectral x-ray imaging was introduced for medical imaging about four decades ago[1,2] and specifically applied to mammography about ten years later.[3] Since then, the most common implementation of spectral mammography has been dual-energy imaging, i.e. spectral imaging with two images at two different energy spectra. Dual-energy imaging has been applied for contrast-enhanced K-edge imaging,[4] to measure the amount of contrast material[5] and energy-weighting to improve the image signal-to-noise ratio.[6,7] It has also been used in unenhanced material decomposition with two basis materials, for example, to improve lesion visibility[3,8] to differentiate between lesion types[9,10] and to measure the volumetric breast density.[11]

Unenhanced material decomposition in dual-energy imaging is done under the assumption that x-ray attenuation is determined by only two independent interaction effects, the photoelectric effect and Compton scattering,[1,2] and adding additional energy levels would therefore not add independent energy information. This assumption is practical because the complexity of an imaging system generally increases with the number of energy levels, and the mentioned interaction effects do contribute the largest cross sections in the mammography energy range. Nevertheless, many current mammography applications of spectral imaging are restricted by only having access to two basis materials. The third most probable interaction effect for breast tissue in the mammography energy range is Rayleigh scattering (coherent scattering). Even though the cross section for Rayleigh scattering is substantially lower than for the photoelectric effect and Compton scattering, the effect might still be used as an independent energy dependency in material decomposition if the imaging system has 1) efficient-enough scatter rejection to treat scattering as attenuation, 2) capability to measure at three different energy spectra, i.e. triple-energy imaging, and 3) high enough signal-to-noise ratio to allow for the effect to be detected despite the relatively low cross section. Adding Rayleigh scattering as an independent energy dependency would enable material decomposition with three basis materials, which has the potential to improve virtually all applications of unenhanced spectral imaging. For instance, skin thickness could be measured and accounted for more accurately in measurements of volumetric breast density, characterization of lesions as benign or malignant could be done more accurately, and the effect of breast thickness variations, which has hampered previous attempts to improve the visibility of malignant lesions, could be accounted for.

---

*E-mail: karl.berggren@mi.physics.kth.se

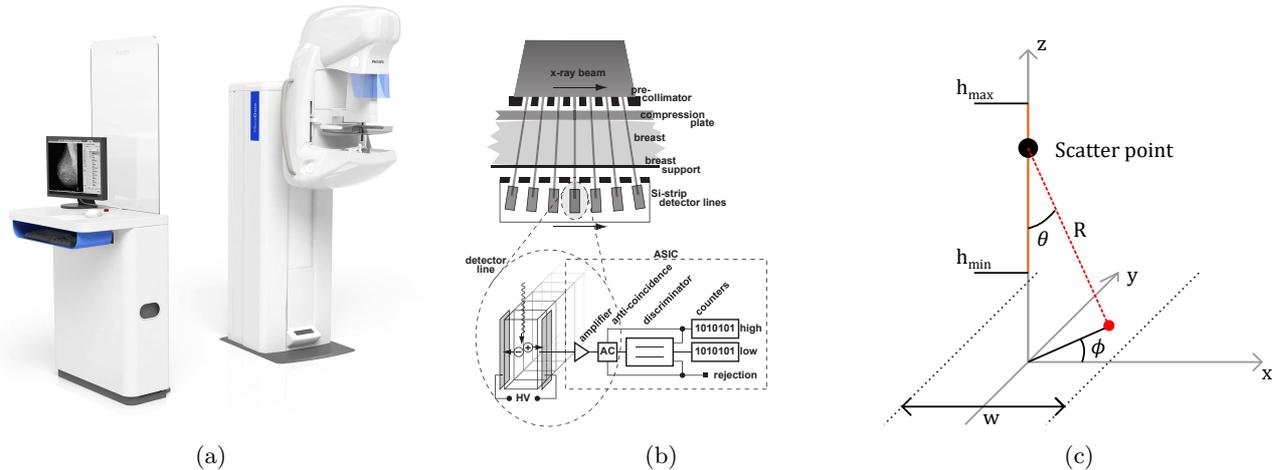

(a) (b) (c)

Figure 1: (a) The Philips Microdose SI system. (b) Schematic of the collimator structure and the photon-counting detector with two energy thresholds. (c) Schematic of a scattering event and the geometry used in Eq. 2, where $x$ and $y$ are the axes on the post-collimator plane, $w$ is the width of a post-collimator slit, $\theta$ and $\phi$ are spherical angles and the object is placed along the $z$-axis between $h_{min}$ and $h_{max}$ above the post-collimator plane.

The purpose of this study is to investigate whether the use of Rayleigh scattering as a third independent interaction effect is feasible on a modified version of a commercially available spectral photon-counting mammography system.

## 2. METHOD

### 2.1 Spectral photon-counting mammography system

The Philips Microdose SI (Philips Healthcare, Solna, Sweden) is a scanning multi-slit mammography system (Fig. 1a and 1b). The image receptor consists of 21 photon-counting line detectors with a pixel pitch of 50 µm.[12] A low-energy threshold in the electronics rejects virtually all electronic noise. Even though the commercially available system features only one high-energy threshold that enables dual-energy imaging, one can simulate a one-shot triple-energy system by performing two image acquisitions with the same exposure settings but with the high threshold at different levels. The main difference between this method and a triple-energy system is the increased exposure with a factor two which is mainly a benefit in a phantom study since it improves image statistics. The main drawback of this implementation in clinical setting is the large risk of vibration and motion blur from the patient which is why in a clinical system one would use a detector system with three or more energy thresholds and acquire one-shot triple-energy images.

The system processes the images from the 21 lines into one image for each energy bin, all detector channels are individually calibrated against PMMA so the output images are equivalent PMMA thicknesses with interchannel variations calibrated away. The three thresholds used produces three different images, the low threshold, that is set just above the electronic noise level, generates a conventional photon-counting image of the whole spectrum referred to as the sum image. The middle threshold generates a medium image of approximately the upper two-thirds of the spectrum and the high threshold generates a high image of the upper one-third of the spectrum.

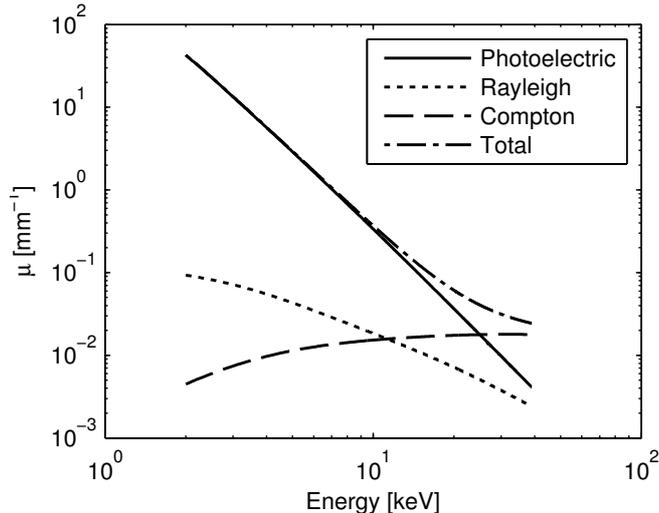

Figure 2: The energy dependencies of the three major interaction effects at mammographic x-ray energies. The plot is logarithmic on both axes to visualize the exponents in the approximation Eq. 1 (the exponent corresponds to the slope of the curve).

## 2.2 Theoretical background

For most natural body constituents at mammographic x-ray energies, it is fair to ignore absorption edges. X-ray attenuation is then made up of three independent interaction effects, namely the photoelectric effect, Compton scattering, and Rayleigh scattering. The contributions of these effects on the linear attenuation, $\mu$, as a function of photon energy, $E$, are approximately

$$\mu(E) = a_{PE} E^{-3} + a_R E^{-1.2} + a_C, \qquad (1)$$

where $a_{PE}$, $a_R$ and $a_C$ are material dependent constants for the respective interaction effects.[13] Equation 1 is an approximation and the actual dependencies are plotted in Fig 2.[14] The photoelectric effect follows an exponential energy dependency quite accurately, but Rayleigh and Compton scattering have somewhat more complex energy dependencies. Nevertheless, the energy dependencies of the three effects differ substantially and any complex behavior can therefore be handled by more advanced calibration models and algorithms.[11,15] For efficient spectral imaging it is necessary to be able to treat scattering processes as attenuation, i.e. scattered photons need to be rejected by the detection system. It has previously been established that the multi-slit geometry of the Philips Microdose SI rejects virtually all Compton scattered radiation[16] but Rayleigh scatted photons have a substantially narrower angular distribution, and the rejection efficiency for these might be lower. The fraction of Rayleigh scattered photons incident on the detector from a line of material with atomic number $Z$, linear attenuation $\mu(E)$, attenuation from Rayleigh scattering $\mu_R(E)$, and located between $h_{min}$ and $h_{max}$ is

$$p_R = \frac{\int_{h_{min}}^{h_{max}} dz \int_{slit} \sin\theta d\theta d\phi \frac{1+\cos^2\theta}{R^2} |F(\frac{E}{hc}\sin\frac{\theta}{2}, Z)|^2 q_0(E) e^{-\mu(E)(h_{max}-z)} \mu_R(E)}{\int_{h_{min}}^{h_{max}} dz \int_{total} \sin\theta d\theta d\phi \frac{1+\cos^2\theta}{R^2} |F(\frac{E}{hc}\sin\frac{\theta}{2}, Z)|^2 q_0(E) e^{-\mu(E)(h_{max}-z)} \mu_R(E)} \qquad (2)$$

where $E$ is the x-ray energies spanning from 0 to $E_{max}$, $x$, $y$, $z$ and $R$, $\theta$, $\phi$ are the coordinate systems depicted in Fig. 1c, *slit* indicates the area covered by a slit, $F(v, Z)$ is the atomic form factor, $h$ is Planck's constant, $c$ is the speed of light and $q_0(E)$ is the incident spectrum.

## 2.3 Experimental verification

We have modified a Philips Microdose SI mammography system to perform triple-energy imaging by performing two image acquisitions with different high-energy thresholds. This generates in total four images: two sum images which should be identical, except for quantum noise, and that images the total spectrum above the electronic noise level, one medium energy image with a threshold set to capture approximately the top 66% of the spectrum and one high energy image to capture approximately the top 33% of the spectrum. All exposure settings were calibrated against a PMMA step-wedge so that the images were represented as equivalent PMMA thicknesses. This calibration also removes non-linear detector response and inter-channel variations. One sum image was discarded and the other three were calibrated against a three material phantom consisting of CIRS (CIRS Inc., Norfolk, VA) glandular and adipose tissue equivalent material and aluminium.

For each pixel the system output is three PMMA thicknesses, $(t_1, t_2, t_3)$, that are mapped against the reference materials glandular tissue, adipose tissue and aluminium, $(t_g, t_a, t_{al})$. The three-material calibration was performed by calculating calibration functions for each reference material using linear interpolation on each individual channel. The reference points were determined from the median over a region-of-interest (ROI) for each combination of material. For all images, a background of 5 mm PMMA and 20 mm of 30% glandular CIRS breast-tissue equivalent material was used to get the total material attenuation in the average mammography range. The calibration points used were the 27 possible combinations of $t_g = 0, 10, 20$ mm, $t_a = 0, 10, 20$ mm and $t_{al} = 100, 300, 500$ µm. Verification measurements were done on a separate image set on the mid-points of each calibration region, i.e. the 8 combinations of $t_g = 5$ and $15$ mm, $t_a = 5$ and $15$ mm, and $t_{al} = 200$ and $400$ µm. Pixels with values outside of the calibration range were flagged and not included in the remaining calculations.

## 3. RESULTS

### 3.1 Theoretical analysis

Figure 3a shows a 38-kVp tungsten mammography spectrum along with the fraction each interaction effect makes up of the linear attenuation for a 55 mm thick compressed breast consisting of 5 mm of skin, 15 mm of glandular tissue and 35 mm of adipose tissue. The contribution by Rayleigh scattering is relatively constant in the relevant energy range and accounts for 11% of the total linear attenuation when weighted with the spectrum. The contributions by the photoelectric effect and Compton scattering are 56% and 33%, respectively.

Figure 4 shows the intensity of Rayleigh scattering over an area of $100 \times 100$ mm$^2$ for an object consisting of 40 mm of carbon (an elemental material was used rather than breast tissue in order to simplify calculation of the form factor). The area of a collimator slit is shown as a black line. The fraction of scattered radiation that hits the detector calculated according to Eq. 2 was 1%, i.e. the scatter rejection was found to be 99%. The amount of scattered radiation hitting adjacent detector lines was not included in this number and the actual efficiency will be slightly lower. Still, we expect the scatter rejection efficiency to be sufficiently high for treating Rayleigh scattering as absorption.

### 3.2 Experimental results

The results from the measurements are presented in Table 1 and Fig. 5. All measurement points except point 8 are within the measurement error from the corresponding true points and even though the errors are quite large, there is no clear overlap between the different measured points. I.e. one can differentiate the material differences in the order of the difference selected for the measurements points: 10 mm glandular or adipose tissue or 200 µm aluminium. The error bars in Fig. 5 show the estimated error of the measurement points (one standard deviation). The relatively large errors suggest that improved statistics would be desirable to reduce uncertainty of the results. However, the measurement points do also show signs of systematic deviations which may be caused by noise in the calibration data, variations between image acquisition, or a sub-optimal implementation of the calibration, for example in the selection of ROIs for determining the calibration values and the measurement values.

For some of the measurement ROIs, a high fraction of pixels were discarded because of having out-of-range values. Noteworthy were the points with the highest total attenuation, points 5, 6 and 7, which had 12%, 22% and 10% of the pixels outside of the calibration range.

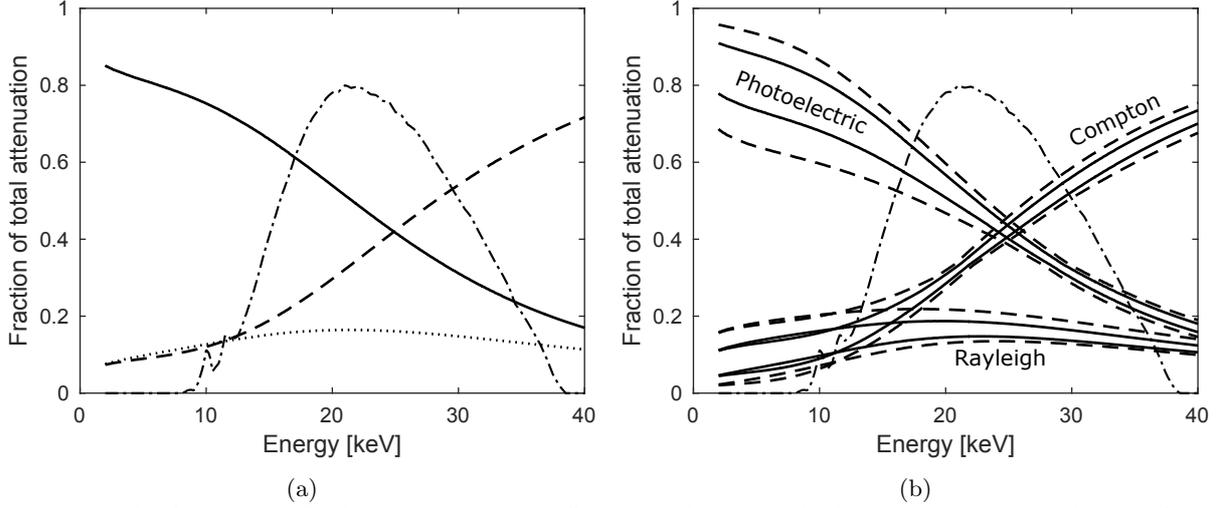

Figure 3: (a) The fraction each photon interaction effect contributes to the linear attenuation. (b) The fractions of attenuation of the total attenuation, similar to Fig. 3a, but with intervals for all measurement and calibration points. The intervals for the measured points are marked out by solid lines and the intervals for the the calibration points are marked out with dashed lines. A 38 kVp tungsten mammography spectrum is included for reference in both plots.

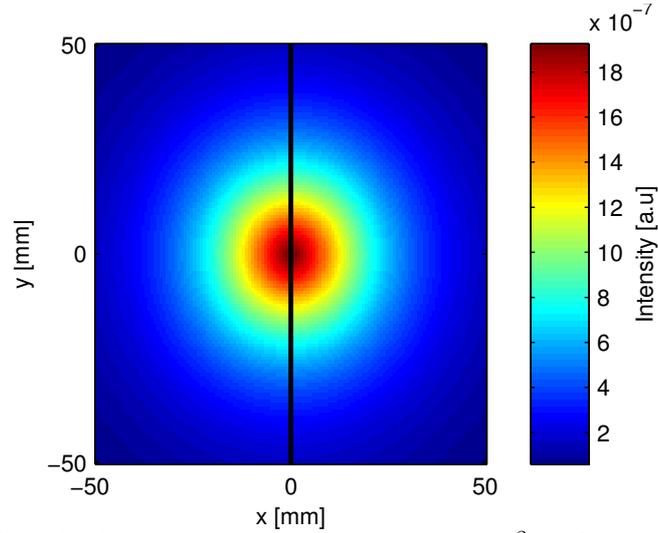

Figure 4: The intensity of Rayleigh scattering onto a $100 \times 100$ mm$^2$ surface. The black line illustrates the width of a post-collimator slit.

| No. | True Adip. | True Gland. | True Al | Meas. Adip. | Meas. Gland. | Meas. Al |
|---|---|---|---|---|---|---|
| 1 | 5  | 5  | 200 | 4.25±3.7  | 5.36±2.96 | 199±76 |
| 2 | 5  | 5  | 400 | 3.54±3.14 | 6.3±2.81  | 364±61 |
| 3 | 15 | 5  | 400 | 12.7±3.15 | 7.16±2.99 | 342±77 |
| 4 | 15 | 5  | 200 | 12.7±3.96 | 7.18±3.65 | 144±91 |
| 5 | 15 | 15 | 200 | 13.5±3.79 | 15.7±2.76 | 168±91 |
| 6 | 15 | 15 | 400 | 12.6±3.73 | 16.3±2.47 | 321±94 |
| 7 | 5  | 15 | 400 | 5.38±3.17 | 14.4±3.04 | 398±61 |
| 8 | 5  | 15 | 200 | 7.77±2.23 | 12.3±2.09 | 260±57 |

Table 1: All measurement points including number references, true values and measured values, including the standard deviation of the measurements.

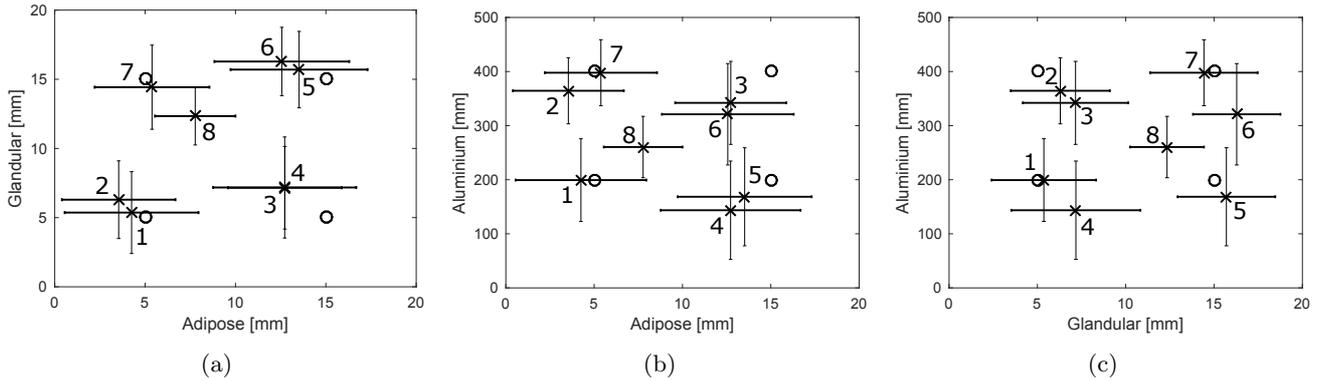

Figure 5: The measured values, marked with x and error bars, together with the true values, marked with circles. The exact values are listd in Table 1. These plots represent the three projections of the 3D volume span by the glandular, adipose and aluminium thicknesses.

## 4. DISCUSSION AND CONCLUSIONS

With this work we set out to show that material decomposition with three basis materials is feasible on a commercially available spectral photon-counting mammography system by testing the following points 1) the scatter rejection of the system is efficient enough that Rayleigh scattering can be treated as absorption, 2) it is conceivable to modify the system to measure at three different energy spectra, i.e. triple-energy imaging, and 3) the signal-to-noise ratio of Rayleigh scattering is high enough to allow for the effect to be useful, at least for large targets.

For point 1) we have theoretically shown that the scatter rejection is high enough to consider Rayleigh scattering as attenuation and thereby introducing a third energy dependence in material attenuation. From the view of the experiment we see that we can decompose images into three materials, but the current experiment is not sufficient to rule out insufficient scatter rejection as a limiting factor. This can however be tested by imaging the same object at different heights above the patient support. This is based on Eq. 2, the scattered photons will have a longer distance to travel before reaching the post-collimator and therefore spread out more and reducing the flux that passes the post-collimator, similar to the inverse square law.

We have succeeded in modifying a commercially available mammography system for triple-energy imaging as in point 2). However, the double exposure method is only suitable for phantom studies and for a clinical implementation a one-shot triple-energy acquisition should be used. Two possible ways of implementing such an acquisition is by alternating the threshold between the different line boards or by using a detector system with three or more electronic thresholds. These methods are however part of ongoing research.

The signal-to-noise ratio is a limiting factor in the current experimental setup and further research is required, both concerning the current implementation to find the limiting factors but also in optimizing other system parameters for improved SNR from Rayleigh scattering. For example, reducing the post-collimator width or increasing the patient support to detector distance would improve the signal from attenuation by Rayleigh scattering.

Improving the electronics and detector is another interesting point. Energy resolution is one possible limitation, photons counted in the wrong bin would increase the noise while the average signal remains correct. Improved handling of charge sharing, chance-coincidence and pile-up are also possible improvements. These questions require further study.

Material decomposition with three basis materials has the potential to improve spectral mammography substantially, for instance, by removing the effects of skin in breast-density measurements and the effects of thickness and density gradients that have so far hampered the development of spectral lesion characterization and detection. The breakthrough work of this study is the implementation of triple-energy imaging on a mammography system and decomposition into three basis materials using attenuation from Rayleigh scattering.

## ACKNOWLEDGMENTS

This work was partly funded by the Swedish Research Council.